\documentclass{JHEP3}
\usepackage{amssymb,amsfonts,amsbsy,amsmath}
\usepackage[ps,dvips,matrix,arrow,frame,import,curve,color]{xy}
\usepackage{epsfig}

\newcommand{\be}{\begin{equation}}
\newcommand{\ee}{\end{equation}}
\newcommand{\ba}{\begin{array}}
\newcommand{\ea}{\end{array}}
\newcommand{\bea}{\begin{eqnarray}}
\newcommand{\eea}{\end{eqnarray}}

\def\half{\frac{1}{2}}

\newcommand{\C}[1]{C^{(#1)}}
\newcommand{\G}[1]{G^{(#1)}}

\newcommand{\Int}{\mathop{\rm Int}\nolimits}

\newcommand{\wdg}{{\scriptscriptstyle \wedge}}

\newcommand{\eq}[1]{Eq.~(\ref{eq:#1})}

\def\tg{{\tilde g}}

\def\tR{{\tilde R}}

\title{Compactification on negatively curved manifolds}

 \author{Michael R. Douglas$^a$\footnote{\mbox{
Email: {\tt douglas@max2.physics.sunysb.edu} }} \,  and  Renata Kallosh
$^{b}$\footnote{\mbox{ Email: {\tt kallosh@stanford.edu} }}
\\

$^a$Simons Center for Geometry and Physics,
Stony Brook University\\
 Stony Brook, NY 11794 USA  \\
\\
I.H.E.S., Le Bois-Marie, Bures-sur-Yvette, 91440 France\\

$^b$Department of Physics, Stanford University, Stanford, CA 94305,
USA }

\abstract{
We show that string/M theory
compactifications to maximally symmetric space-times
using manifolds whose scalar curvature is everywhere
negative, must have significant warping, large stringy corrections, or both.
}

\begin{document}

\section{Introduction}

If our universe is described by string/M theory, there exist six or seven extra dimensions
of space, not yet detected by experiment.  This is possible if the extra dimensions
take the form of a small, compact manifold $M$.  There are many possible choices of
$M$ and it is very interesting to find constraints on this choice, both internal consistency
conditions, and those required to match the observations.

In this note we find constraints on the possibility of using a $k$-dimensional manifold
$M^k$ with a metric of negative curvature.  Mathematically, the
characterization of a manifold with metric as having positive, zero, or negative
curvature is one of the most important aspects of its geometry  \cite{Berger}.
Familiar examples of the first two classes are the sphere $S^k$ and the torus $T^k$, while
the higher genus Riemann surfaces provide an infinite set of
negatively curved examples in two dimensions.

Examples of negatively curved manifolds with $k>2$ include hyperbolic manifolds,
high degree hypersurfaces, nilmanifolds and solmanifolds in $k=3$,
products of lower dimensional negatively curved manifolds, and others.
While these may be less familiar, and
since the curvature tensor has many components even the definition of ``negative
curvature'' requires discussion,
there are in a precise sense (as explained in  \cite{Berger}) many more
negatively curved than non-negatively curved manifolds, just as for $k=2$.

Using this trichotomy, we get a basic (highly oversimplified) classification of supergravity
and string theory compactifications.
The simplest case is zero curvature, here meaning Ricci flat.
In this case, $M$ solves the vacuum Einstein equations, and we get a compactification to
Minkowski space-time.
This includes tori, Calabi-Yau manifolds, $G_2$ manifolds,
and singular limits of these such
as orbifolds.  One can of course turn
on more fields and vary the Ricci curvature away from zero, but the global features of the solution will
not change, unless we leave the supergravity regime.

For many manifolds, no Ricci flat metric is believed to exist.
In this case, one must try to solve the Einstein equation with a source of stress-energy,
say from flux ($p$-form gauge fields).  In the simplest and by now very well-known
solutions of this type, space-time is a direct product of anti de-Sitter space-time and a
sphere.  Thus, the internal manifold has positive curvature.  This class of solution
can be generalized to many other positive curvature manifolds, obtained by quotienting
by a group action, blow-up, and various other constructions.

Both constructions lead to a wide variety of compactifications on $M$ with positive or zero
Ricci curvature.
On the other hand, there are few known families of solutions based on other types
of $M$, with negative or mixed Ricci curvature.  Indeed, at this writing, there are no such 
compactifications to maximally symmetric space-times 
(Minkowski, de Sitter or anti-de Sitter) for which the
higher dimensional field configurations are known explicitly or at least proven to exist.
Assuming this can be done, we would like to decide whether such solutions are common or else rare,
relying on structures and coincidences which only hold in a few examples.

One particular reason to be interested in this is to find de Sitter solutions. It was noticed in \cite{Silverstein:2007ac} that  the four dimensional curvature gets a significant positive contribution from the negative curvature of the internal space.
This could serve as an important contribution to the uplifting of the string theory vacua which tend to be anti de-Sitter solutions, rather than de Sitter.
Indeed,  we will show, starting with ten dimensional Einstein equations, that the four dimensional curvature scalar $R_4$ is defined by the
four dimensional energy-momentum tensor trace $T_4$ as well as by the  six dimensional curvature scalar  $R_6$
\be \label{eq:basic2}
 R_4 = - T_4 - 2 R_6
\ee
Clearly, a  negative curvature of the compactified manifold $R_6<0$
may, in principle,  overcompensate the possible negative contribution from $T_4$ term and produce a de Sitter minimum with $R_4>0$. This feature of the negative curvature has allowed to present an explicit class of de Sitter vacua in massive type IIA supergravity \cite{Romans:1985tz} with local sources, which evade the no-go theorems discussed in \cite{DeWolfe:2005uu}. The analogous situation concerns the no-go theorems for slow-roll inflation in \cite{Hertzberg:2007wc}.  The presence of the negative curvature in model \cite{Silverstein:2008sg} helps to provide a slow-roll inflationary model which makes an interesting prediction of a detectable level of gravity waves.
See \cite{Greene:2010ch} for another recent work linking negative curvature and cosmology.

In this note, we study maximally symmetric 
supergravity solutions.  There
is a very simple ``no-go'' argument, which we give in section 2, 
that in the absence of warping,
and if the stress-tensor satisfies an energy condition which holds for all the standard
forms of matter (including the mass term of IIa supergravity),
the internal manifold $M$ must have nonnegative scalar curvature at each point.
While both conditions can be relaxed in string/M theory compactification, this simple argument
provides a useful starting point for a more general discussion.

The argument is very similar to the no-go theorem of
\cite{Gibbons-review,de Wit:1986xg,Maldacena:2000mw}.
We identify the combination of
the Einstein equations which determines the scalar curvature, and find an inequality
satisfied by the source.  However an important difference in the present case is that this
imposes a condition at each point of $M$.  As we discuss in section \S \ref{ss:localized},
most of the string/M theory terms which
violate this inequality, such as orientifold planes, are {\bf localized}.  Thus, there will
be many regions in $M$ in which they do not contribute, and thus they cannot by
themselves evade this ``no-go'' result.

In section 3, we discuss ways to get negative curvature which could work within 
the regime described by supergravity with branes and stringy corrections.
Of course, we know of string/M theory solutions which are not well described by supergravity,
but since the concept of negative curvature (or even curvature) has not yet been defined for
such solutions, it is too early to place them in the general scheme we are discussing.

In section 4, we comment on the possible implications of this for solutions which
have been shown to exist using effective potential arguments.

\section{Basic argument}

We focus on string theory and $d=10$ supergravity with local sources.
In M-theory and $d=11$ supergravity with local sources the results are analogous. Our assumptions are:
\begin{enumerate}
\item We assume that 10d is a product of 6d and 4d and the 4d space is symmetric, {\it i. e. }
$T_{\mu\nu} \sim g_{\mu\nu}$ and form fields $F_{\mu\nu\rho\sigma mn...} $ are proportional to $\epsilon_{\mu\nu\rho\sigma}$.
\item We consider the simplest case that the geometry is not warped: the function of 6d 
$y$ coordinates denoted $\Delta^{-1}(y) $ in \cite{de Wit:1986xg} and $\Omega^2(y) $ in \cite{Maldacena:2000mw}  is  constant, $\Delta(y) \sim \Omega (y) \sim \rm const$.
\end{enumerate}

Starting from the Einstein equations,
\be
 R_{MN}-\half g_{MN} R=  T_{MN}
\ee
we pass to the trace reversed form,
\be \label{eq:einst}
R_{MN}= T_{MN} - \frac{1}{8} g_{MN} T^L{}_L .
\ee
Here $L=0, ..., 9$, $m=1,...,6$ and $\mu=0,..,3$.
We introduce the notation
\be
R_4= R_{\mu\nu} g^{\mu\nu} \ , \qquad R_6= R_{mn} g^{mn}\ ,
 \qquad T_4= T^\mu {}_\mu \ ,\qquad T_6= T^m{}_m
\ee
If there is no warping, so the metric is a direct product, the curvature $R_4$ is
constant on $M$, and is zero, positive or negative for Minkowski, dS and AdS
space-times respectively.

By tracing \eq{einst} we find that
\be\label{eq:trace4}
\boxed{
R_{4}= - \half T_6 + \half T_4}
\ee
and
\be\label{eq:trace6}
\boxed{
R_{6}= + \frac{1}{4} T_6 - \frac{3}{4}  T_4} .
\ee
Note that combining \eq{trace4} with \eq{trace6} multiplied by 2, we reproduce \eq{basic2}.
The first of these
coincides with eq. (33) in \cite{Maldacena:2000mw} and (2.12) in \cite{de Wit:1986xg} (taking account of
their conventions for the curvature), comparing in both cases with the non-warped case.
The statement that its right hand side is non-positive
is called the strong energy condition
\cite{Wald:1984rg,Gibbons-review}.
It is satisfied by all standard forms
of matter except for negative potential energy, and ``$0$-form'' flux \cite{Romans:1985tz}.  Thus, it
is very easy to see that de Sitter space-time, with $R_4>0$, cannot be obtained without either
sources which violate the strong energy condition, or warping.

With a little more work, one can show that warping does not affect this conclusion.
Although it makes another
contribution to \eq{trace4}, after integrating the equation over $M$ with a suitable
positive weight, the extra contribution vanishes.

Clearly we can apply the same reasoning to \eq{trace6}.
Generalizing $4$ and $6$ to arbitrary dimensions $d$ and $k$, the claim is
\be \label{eq:claim}
0 \le R_k = \frac{d-2}{d+k-2} T_k - \frac{k}{d+k-2}  T_d.
\ee
Let us check that its right hand side is non-negative for
the standard forms of matter, now including $0$-form flux.

The only input we need from supergravity is that the stress-tensors for the form
fields are quadratic in the field strengths, and positive.  In general, this is not completely
obvious, because of the presence of Chern-Simons terms and modified Bianchi identities.
However there is a formalism in which it is obvious, the
democratic version of $d=10$ supergravity \cite{Bergshoeff:2001pv}, as we review
in an appendix.

\subsubsection{Magnetic $p$-form flux}
\label{sss:mag}

 \be \label{eq:mag-six}
 T_{mn}^{mag}=
  p F_{m..} F_n^{..} - {1\over 2}g_{mn} F^2_p \qquad \Rightarrow T_6^{magn-p-form}= (p-3) F^2_p
 \ee
so
 \be \label{eq:mag-four}
 T_{\mu\nu}^{mag}=-  {1\over 2} g_{\mu\nu} F^2_p \qquad \Rightarrow T_4^{magn-p-form}= -  2 F^2_p
 \ee
Evaluating our general  \eq{trace4} and \eq{trace6} we find
 \be \boxed{
R_{4}^{mag}= - {1\over 2} T_6 + {1\over 2} T_4=   {1-p\over 2} F^2_p}
\ee
and \be \boxed{
R_{6}^{mag}= + {1\over 4} T_6 - {3\over 4}  T_4 = {p+3\over 4} F^2_p} ,
\ee
verifying the claim.  In $d$ and $k$ dimensions, the second equation becomes
\be \label{eq:total-Rk-source}
R_{k}^{mag}= \frac{1}{d+k-2} \left( (d-2)T_k - k T_d\right) =
 \frac{ (d - 2 )p +  k}{d+k-2}  F^2_p .
\ee

Another way to think about this result is to note that the $k$-dimensional trace of the
Einstein equations is proportional to the variation of the action with respect to the
$k$-dimensional conformal factor (say as defined below in \eq{def-conformal}).
Integrating over $M$, one relates this to the variation of the effective potential with respect to
the volume of $M$.  Writing this volume in terms of a length scale $L$ as,
\be
\int_M \sqrt{g} \equiv L^k ,
\ee
naively this is
\bea
L\frac{\partial}{\partial L} &\int \sqrt{g} g^{i_1j_1}\cdots g^{i_pj_p} F_{i_1\ldots i_p} F_{j_1\ldots j_p} \\
& \sim (k-2p) L^{k-2p} ,
\eea
as in \eq{mag-six}.  This term can make contributions of either sign to $R_6$.

However, there is an additional factor from the rescaling to go to $d$-dimensional
Einstein frame.
Using the results of \cite{Douglas-warp}, one can show that this is
\be
\sqrt{g^{(d)}} = u^2 \sim L^{-kd/(d-2)} ,
\ee
which corresponds precisely to the $T_d$ dependence  in \eq{total-Rk-source}.

 \subsubsection{Electric flux}

It is not hard to see that an electric contribution is dual to a magnetic flux,
with $p$ replaced by $10-p=\tilde p$, for example,
\be
R_{6}^{el-p-form}= {13-p\over 4} \tilde f^2_p =  {\tilde p+3 \over 4} \tilde f^2_{10-\tilde p} \qquad  \Leftrightarrow \qquad R_{6}^{magn-p-form}=  {p+3\over 4} F^2_p
\ee
Let us consider an electric form field with $p\geq 4$, with 4 legs in space-time and the rest in 6, i. e. for
$F_{\mu\nu\rho\sigma m_1... m_{p-4}}= i e \epsilon_{\mu\nu\rho\sigma} f_{m_1... m_{p-4}}$ as shown in \cite{de Wit:1986xg} we get
\be
T_{mn}^{el-p-form}= -p(p-1)(p-2)(p-3)\left [  (p-4)f_{mq_{1}...q_{p-5}} f_n^{q_{1}...q_{p-5}} -{1\over 2} g_{mn} f_{q_1...q_{p-4}} f^{q_1...q_{p-4}} \right ]
\ee
\be
T_{\mu\nu}^{el-p-form}= -{1\over 2} g_{\mu\nu}  \tilde f^2_p
\ee
where
\be
p(p-1)(p-2)(p-3) f_{q_1...q_{p-4}} f^{q_1...q_{p-4}}\equiv \tilde f^2_p
\ee
and
\be
T_{6}^{el-p-form}=  (7-p) \tilde f^2_p
\ee
\be
T_{4}^{el-p-form}= -2   \tilde f^2_p
\ee
After substituting $p\rightarrow 10-p$,
these agree with Eqs. (\ref{eq:mag-six}) and (\ref{eq:mag-four}).  The same is true
for the linear combinations $R_4$ and $R_6$.

\subsection{Localized sources}
\label{ss:localized}

Following \cite{Giddings:2001yu}, consider a p-brane wrapped on a (p-3)-cycle $\Sigma$
\be
S_{loc} = - T_p \int_{R^4\times \Sigma} d^{p+1} \xi \sqrt{-\tilde g} + \mu_p \int_{R^4\times \Sigma} C_{p+1}
\ee
where $T_p= |\mu_p| e^{p-3 \phi/4}$ and $g$ is the determinant of the metric induced on the brane, $\tilde g_{\alpha \beta} = {\partial X^M\over \partial \xi_{\alpha}}  {\partial X^N\over \partial \xi_{\beta}} g_{MN}$. When the brane is embedded into a 4 space-time dimensions and p-3 internal dimensions, i.e. $X^\mu=\xi^\mu$ and $X^{i}= \xi^i, i= 1,...p-3$, the variation of the Born-Infeld action leads to
\be
T_{\mu\nu}^{loc}= -T_{p} g_{\mu\nu} \delta (\Sigma) ,  \qquad T_{mn}^{loc} = -T_p \Pi_{mn}^\Sigma \delta (\Sigma) ,  \qquad \Pi_{mn}^\Sigma g^{mn}= p-3
\ee
Here $\delta (\Sigma) $ means that the energy-momentum tensor from local sources is concentrated only at the (p-3)-cycle $\Sigma$. The projector  $\Pi_{mn}^\Sigma$ from the 6d space into a (p-3)-cycle shows that the brane action does not depend on the components of the metric  in the direction orthogonal to the cycle, $i'= (p-4), ...,6$.

So we see that
\be
T_4^{loc}= -4 T_p \delta (\Sigma) \qquad T_6^{loc}= - T_p (p-3) \delta (\Sigma)
\ee

 \be\boxed{
R_{4}^{loc}= - {1\over 2} T_6 + {1\over 2} T_4={p-7\over 2} T_p \delta (\Sigma)}
\ee
and \be \boxed{
R_{6}^{loc}= + {1\over 4} T_6 - {3\over 4}  T_4 = {15-p \over 4} T_p \delta (\Sigma)}
\ee

We again find that these sources satisfy the expected inequalities, as long as the
brane tension $T_p$ is positive.  Following the discussion at the end of \S \ref{sss:mag},
this will be true for any source with positive energy which falls off as the volume of $M$
goes to infinity.  As pointed out in \cite{Giddings:2003zw}, this falloff would fail to hold
only for an energy density which grows at least as fast as the volume, which seems highly
unphysical.  Thus the only way around the inequality is to use sources with $T_p<0$.

Of course, string theory contains such sources, for example
orientifolds.  Can they change the conclusion?
As we noted in the introduction, the inequality \eq{claim} must hold at {\bf every point}
in the internal manifold $M$.  On the other hand, most branes and orientifolds are
lower dimensional objects, and thus contribute only on submanifolds of $M$.  Of course
there is some sense in which these contributions are spread out over the string and
Planck scales, but to analyze the solutions in detail so close to the source, we must leave
the supergravity approximation, instead relying on open strings, gauge theory, or some other
description.  To work purely within supergravity, one
normally one treats branes as delta function sources; we see
no reason this approximation should be wrong in general (of course there might be subtleties
in particular cases).  Thus, even including branes and orientifolds supported on lower dimensional cycles,
the inequality \eq{claim} still leads to significant constraints, and rules out solutions with scalar
curvature everywhere negative.

In principle, the O9 might provide a counterexample, but anomaly cancellation will
force there to be sufficiently many D9 branes to cancel this source.

Another idea sometimes called upon in this context is to spread out or ``smear'' an orientifold
or brane source \cite{Acharya:2006ne}.
It was applied in the context of the  IIA flux vacua \cite{DeWolfe:2005uu} of massive 10d
supergravity \cite{Romans:1985tz} with local sources.

The idea is to replace a localized source by a smooth source with support throughout $M$, whose
integral over $M$ is the same as the original source.  While this certainly simplifies the
task of finding a solution, and would avoid the issue we are describing,
in the context of string compactification to dimensions $d>2$, it is not justified.  First, in
classical (genus zero) string theory, the location of a brane or orientifold plane is a modulus,
which must be fixed to define the theory.  Of course, in the full quantum theory one can propose
a wave function which is delocalized in the moduli, and in this way define states in which the
expectation value of the metric and other fields is smeared in this way.  However, in dimensions
$d>2$ this amounts to combining different superselection sectors, and such an
expectation value is not physical; in a given superselection sector the expectation value of a
brane or orientifold position will take a fixed value.  This is not particular to string/M theory but
is a general feature of quantum scalar fields in dimensions $d>2$.

Smearing is valid when constructing solitonic particle or string solutions, because their wave functions
can superpose different values of the zero modes.  Even in $d>2$, the
question of whether smeared solutions exist is not without interest, as if they do not,
one expects that no localized solution will exist either.  However it is not
really an answer to the issue we are raising.

\subsection{Total curvature, no warping case}

Combining the previous results,  the conclusion is that
 \be\boxed{
R_{4}=  - {p-1\over 2} F^2_p +{p-9\over 2} \tilde f^2_p +{p-7\over 2} T_p \delta (\Sigma)}
\label{eq:4dcurv}\ee
and \be
\boxed{
R_{6}= {p+3\over 4} F^2_p +{13-p\over 4} \tilde f^2_p +{15-p \over 4} T_p \delta (\Sigma)}
\label{eq:6dcurv}\ee

The second term in each eq. is valid only for $p\geq 4$, so it is negative in $R_4$ and positive in $R_6$. In the third term $p\geq 3$ and the sign of $T_p$ is positive for the D-p-branes and negative for the O-p-planes. It is nice to see that in both equations the duality condition with replacement of $p$ into $10-p$ and electric into magnetic fluxes, works in agreement with the generic condition given in \eq{duality}. Namely, in \eq{4dcurv} as well as in \eq{6dcurv}
\be
p\rightarrow 10-p \, ,  \qquad F_p^2 \rightarrow \tilde f^2_p
\ee
is a symmetry of the flux contributions to the corresponding curvatures.

It may be also useful to provide the energy-momentum tensor sub-traces.
 \be
T_{4}=  - 2 F^2_p -2   \tilde f^2_p -4 T_p \delta (\Sigma)
\ee
and \be
T_{6}= (p-3) F^2_p +(7-p) \tilde f^2_p +(3-p) T_p \delta (\Sigma)
\ee
One can verify that $R=R_4+R_6$ and $T=T_4+T_6$ are related to each other by a factor $-4R=T$ for each of the 3 contributions.

By looking at \eq{6dcurv} we see that fluxes always give  positive contributions.
While local sources may give a negative contribution, their negative energy is localized on cycles on which the branes are wrapped.
Thus, using the sources discussed so far, there can be no solution of the 6d Einstein equations
with everywhere negative 6d  curvature.

\subsection{Internal energy conditions}

The results we just obtained suggest that we introduce a new `` internal energy condition''
(or IEC),
\be
T_k \ge \frac{k}{d-2}  T_d ,
\ee
which looks very similar to the strong energy condition (SEC),
\be
T_k \ge \frac{k-2}{d}  T_d .
\ee
Their relationship depends on the sign of $T_d$.  Since
$T_d = -d\Lambda_{matter} = -d\rho$, this will normally be negative,
in which case the IEC follows from the SEC and is strictly weaker.

We could also consider a
``strong internal energy condition,'' which
would guarantee that $R_{ij}$ be non-negative definite.\footnote{
This idea was discussed in passing in \cite{Acharya:2006zw}.}
Everywhere non-negative Ricci curvature is a far stronger geometric constraint
than non-negative scalar curvature, ruling out the vast majority of manifolds \cite{Berger}.
This would hold if
\be
T_{ij} e^i e^j \ge \frac{1}{D-2} (T_d + T_k) |e|^2 \qquad \forall e\in TM .
\ee
Is this ever violated?  For magnetic fluxes, this becomes
\be \label{eq:mag-bound}
|e^m F_{m..}|^2 \ge \frac{p-1}{p(k+d-2)}  F^2_p |e|^2 \qquad \forall e\in TM .
\ee
For several fluxes, one would have a single condition obtained by adding the left
and right hand sides separately.  Note also that under $p\rightarrow k-p$ and
$F\rightarrow *^{(k)}F$, the condition changes, but in a rather minor way; $(p-1)/p$
becomes $(k-p-1)/(k-p)$.

Intuitively, this condition will hold at a point if the flux is not too anisotropic,
meaning that the ratio between flux in different planes stays below $(k+d-2)$.
It  is clearly true for $p=k$.  And, since the coefficient on the right hand side is small,
one might expect it to ``usually'' hold.  But there are counterexamples, with the simplest
perhaps being the supergravity solutions of \cite{Maldacena:2000mw}.  These are
obtained by compactifying some of the spatial AdS dimensions, and are not maximally
symmetric in the remaining space-time dimensions, but they illustrate the possibility.

For $p=2$ and $k-p=2$, which are the cases of most interest in IIa compactification
with $k=6$, one can understand this condition
using a change of basis to put $F$ or $*F$ in canonical form,
which is $F_{2n-1,2n}=-F_{2n,2n-1}$, all other components zero.  In this case \eq{mag-bound}
becomes
\be
F_{2n-1,2n}^2 \ge \frac{p-1}{p(k+d-2)} \sum_n F_{2n-1,2n}^2 .
\ee
While clearly this can be violated, doing so requires the flux to be quite anisotropic.

\section{Stringy corrections, conformal and warp factor dependence}

\eq{6dcurv}  shows that almost all sources to $R_6$ are positive; the only negative
sources in IIa compactification in the supergravity limit are the O6 planes, which are localized.
Thus we cannot find solutions with everywhere negative scalar curvature, as is the case
for the nilmanifolds used in \cite{Silverstein:2007ac}.

There are several ways one can try to get around this in string/M theory.  One is to call upon
higher derivative terms or quantum corrections.  Another is to vary the metric.  Although this
possibility is too difficult to analyze in generality, one can make fairly simple statements about
the dependence on the warp and conformal factors.

\subsection{Stringy corrections}
\label{ss:stringy}

The higher derivative terms of string/M theory,
such as the $R^4$ term, see for example \cite{Becker:2002nn}, clearly are not localized, and
need not satisfy the positivity condition \eq{claim}.  In fact, these terms
can sometimes be shown to give negative contributions.
For example, the $C^3 \wedge R^4$ term of M theory, in the context of compactification
on a Calabi-Yau fourfold as in \cite{Becker:1996gj}, is topological, in fact the curvature term integrates
to the Euler characteristic.

Of course, from the point of view of supergravity, these terms are corrections which are
suppressed by factors such as the curvature of $M$ in string units.  On the other hand,
the fluxes are also quantized in string units, so that the leading stringy
corrections can come in at the same order.  This is familiar in the case of tadpole conditions,
which relate topological numbers such as the number of units of flux, numbers of branes,
and the Euler character of $M$.

In supersymmetric solutions, the Einstein equations we are discussing are usually partnered
with tadpole conditions, making it natural to get Minkowski space-time.  Without supersymmetry,
this need not be the case.  What sort of negative curvature can a term like this support?
Let us write the 6d trace of the Einstein equations schematically as
\be
R_6 \sim -l_s^6 ({\rm Riemann_6})^4 + T_6 + \ldots .
\ee
The fourth order term is written this way to emphasize that it depends on all
components of the Riemann tensor, and can be nonzero even if $M$ is Ricci flat.
More generally, the magnitude of the Riemann tensor, and that of
the scalar curvature $R_6$, are {\it a priori} not the same.

Assuming the negative $R^4$ term dominates, and estimating ${\rm Riemann_6} \sim 1/L^2$
where $L$ is some characteristic size of $M$ (say the diameter), we can have
$R_6 \sim l_s^6/L^8$, with further corrections suppressed by $l_s^2/L^2 << 1$.
This is the same scaling as $p=3$-form flux.  
 The string theory $R^4$ corrections of the type studied in \cite{Becker:2002nn} have been used in  cosmology  with large-volume flux compactifications in
\cite{Balasubramanian:2005zx}. It remains to be seen if the analogous string type corrections may be useful for supporting the negative curvature of the the Nil 3-manifolds of \cite{Silverstein:2007ac}, \cite{Silverstein:2008sg}.


\subsection{Warp factor}

 Starting with a maximally symmetric space-time metric $\eta_{\mu\nu}$
and an internal metric $g_{ij}$, we consider the metric ansatz
\be
\label{eq:metric}
ds^2 \equiv \tilde g_{MN} dx^M dx^N = e^{2A(y)}\eta_{\mu\nu} dx^\mu dx^\nu +
 e^{2B(y)} g_{ij} dy^i dy^j ,
 \ee
with two freely chosen functions $A$ and $B$.  Simple computations done in
\cite{Douglas-warp} allow generalizing \eq{4dcurv}  and  \eq{6dcurv}
to this case.

This modifies both equations, and
the modified $R_6$ equation can be found in equation (5.2) in \cite{Douglas-warp},
\be \label{eq:k-trace-eqn}
\frac{k-2}{2} R^{(k)} = 2(k-1) u^{-1} \nabla^2 u + \frac{2(D-2)}{d}(u^{-1}\nabla u)^2
 -\frac{k}{2} u^{-4/d} C -T^{(k)} , 
\ee
where $u=e^{dA/2}$.
This has a sufficiently complicated dependence on the warp factor to
counter the no-go theorem; in particular
the $u^{-1} \nabla^2 u$ term can easily be negative.

Of course, one must then ask whether there is some analog of the arguments in
\cite{Gibbons-review,de Wit:1986xg,Maldacena:2000mw}
which allowed rewriting these terms as a total derivative, thus getting a constraint on
the integral, independent of warping.  Thus, let
us consider the most general integral constraint, obtained by integrating \eq{k-trace-eqn}
against a weight $u^\alpha$, and adding $\beta$ times the warp factor constraint
(Eq. (2.33) in \cite{Douglas-warp}),
\be \label{eq:u-constraint}
-\frac{d-2}{2}R^{(d)} u^{1-4/d} = -{2(d-1)} \nabla^2 u +\left(\frac{d}{2}R^{(k)}+T^{(d)}\right)u ,
\ee
integrated against the weight $u^{\alpha-1}$.  
Integrating by parts and multiplying by $2$, this leads to
\bea
(k-2-\beta d)\int u^\alpha R^{(k)} &=&
\left(\frac{4(d+k-2)}{d}-4(\alpha-1) (k-1-\beta(d-1)\right) \int u^{\alpha-2} (\nabla u)^2
 \nonumber\\
&&+(\beta(d-2)-k) \int  u^{\alpha-4/d} R^{(d)}
+ \int u^\alpha \left( 2\beta  T^{(d)} -2 T^{(k)} \right) .
\eea

To get the combination in \eq{claim}, we take $\beta=k/(d-2)$.
Multiplying by $-(d-2)/2(d+k-2)$, we obtain
\be \label{eq:int-constraint}
\int u^\alpha R^{(k)} =
-2\left(\alpha-\frac{2}{d}\right) \int u^{\alpha-2} (\nabla u)^2 
+\frac{1}{d+k-2}\int u^\alpha \left( (d-2)T^{(k)} - {k} T^{(d)}   \right) .
\ee
Taking $\alpha=2/d$, the term involving derivatives of the warp factor
drops out, to give an integral version of \eq{claim} directly analogous to that in
the no go theorem of \cite{Gibbons-review,de Wit:1986xg,Maldacena:2000mw}.
Thus, warping alone will not allow a supergravity solution with everywhere negative
curvature.  On the other hand, this integrated type of no go theorem can be violated
in string/M theory by localized sources such as O6 planes.

To get a condition which applies at each point of $M$,
one can find another linear combination
of the $R_4$ and $R_6$ equations in which the warping contribution is non-positive definite.
This is equation (5.6) in \cite{Douglas-warp}, which in these dimensions is
\be
R_6 = -12 (\nabla A)^2 +  e^{-2A} R_4 +\frac{3}{4} T_6 - \frac{5}{4} T_4 .
\ee
While the combination of $T_6$ and $T_4$ appearing
here is different from that in \eq{trace6}, one can check that it too is
always positive for fluxes.  Thus, one sees that, at least for de Sitter and Minkowski
space-times, the same conclusion would apply (ruling out negative scalar curvature)
without warping, but warping provides a negative contribution to the scalar curvature.

Physically, by the discussion in section 3 of \cite{Douglas-warp},  the total vacuum energy
is an integral of the local energy densities weighted by the warp factor.  Since the warp factor
is large in regions of negative curvature and positive energy, and small in regions of
positive curvature and negative energy, this weighing will also tend to increase the total
vacuum energy and thus favor de Sitter.  This weighing (by $u^2$) is different from that in 
\eq{int-constraint}, so it is still possible that negative curvature is relevant for this.

\subsection{Conformal factor}

We now consider an internal metric
\be \label{eq:def-conformal}
\tg_{ij} = e^{2B} g_{ij} ,
\ee
where $B$ is a function on the compactification manifold.  The formula expressing
the curvature $\tR_6$ of $\tg$, in terms of $R_6$ and $B$ , is standard (e.g. see the
appendix of Wald's textbook):
\be
\tR_6 = e^{-2B} \left( R_6 -10\nabla^2 B  -20(\nabla B)^2 \right).
\ee
Locally, this can modify $R_6$ in either direction (more positive or more negative).

Since the extra terms are not a total derivative,
there is no topological constraint on the integrated scalar
curvature (in greater than two dimensions), and varying the conformal factor
will change the integrated scalar curvature.
The overall change will tend to be positive.
This can be seen by computing
\bea
\int \sqrt{\tg} \tR_6 &=& \int \sqrt{g} e^{4B} \left( R_6 -10\nabla^2 B  -20(\nabla B)^2 \right) \\
&=& \int \sqrt{g} e^{4B} \left( R_6 + 20(\nabla B)^2 \right)
\eea
after an integration by parts.  This conclusion can be strengthened by making the change
of variables
\be
v = e^{2B} ,
\ee
in terms of which
\be
\int \sqrt{\tg} \tR_6 = \int \sqrt{g} \left( 5(\nabla v)^2 + v^2 R_6 \right).
\ee
We can regard this as a functional of the conformal factor $v$, then
there is a global constraint -- if we set
the overall scale of the metric, say by constraining the total volume, then
the integrated scalar curvature has a definite minimum value.
This is the subject of the Yamabe problem in mathematics \cite{yamabe}.
The minimum tends to be realized by the constant curvature metric,
while a nonconstant conformal
factor will increase the integrated scalar curvature.

There are topological constraints on the manifolds which
can realize metrics with positive scalar curvature at every point,
and even on those which can have non-negative scalar curvature at every point
\cite{Berger}.  However, if we weaken this to allow negative scalar curvature at even a single point,
in $k\ge 3$ there is no constraint \cite{KazdanWarner}.  A more physics-friendly
explanation of this point can be found in \cite{McInnes:2007pg}.

The conformal factor is used in well known solutions such as \cite{Becker:1996gj} and
the GKP vacua \cite{Giddings:2001yu}.  These are flux compactifications on Calabi-Yau manifolds,
and one might worry about the same type of contradiction: the flux is a positive source
to \eq{6dcurv}, but of course a Calabi-Yau manifold has $R_6=0$.  The resolution
is that these metrics are only conformally Calabi-Yau.

Taking this into account, we can imagine finding a IIa solution in which the curvature is negative
near the O6 planes, and positive elsewhere.
A good example is the IIa solution of \cite{Kachru:2002sk}, which is obtained
by T-duality from a GKP flux solution on a torus. 
Of course, since this will increase the total integrated scalar
curvature, this would be expected to
reduce or eliminate the effect of negative curvature in favoring de Sitter
space-time.  
In the Calabi-Yau solutions, arguments using supersymmetry and no-scale
structure showed that
the solutions remained Minkowski (at least, before fixing K\"ahler moduli and uplifting).
At present we have no analog of these arguments for de Sitter solutions.

\section{Discussion}

Except for a few very symmetric examples,
finding explicit compactifications of supergravity by solving the equations of motion is
very difficult.  After a great deal of study, there are more or less two successful approaches.

One is to appeal to mathematical existence theorems showing that certain solutions
exist, and then perturb around these.  This is how the original Calabi-Yau compactifications
of the heterotic string were justified, and barring the discovery of exact metrics on Calabi-Yau
or other manifolds leading to quasi-realistic compactifications,
is likely to remain the ``gold standard.''  From this point of view, further development
of a more mathematical approach to string compactification, 
as illustrated for example by \cite{Fu:2006vj}, would be very helpful.

The other approach is to assume that the problem is governed by a lower dimensional effective theory,
construct its effective potential, and argue that it has a local minimum.  In fact one can think of
the solution of supergravity equations of motion as minimizing an effective potential 
\cite{Giddings:2005ff,Douglas-warp},
so the new assumption here is that one can reduce this to minimizing with respect to
a finite number of fields.  The justification for this \cite{Witten:1985bz} is that one can solve the equations
of motion for massive fields in a perturbative series, schematically
\be
M^2 \phi^i = J^i - g^i_{jk} \phi^j \phi^k + \ldots .
\ee
The meaning of this expression is as follows.
Suppose one has a ``near-solution'' which approximately solves the equations of motion;
one can try to develop a Kaluza-Klein expansion around it, in which variations of the massive fields
are denoted $\phi^i$.  Since it is not a solution, the sources $J^i$ will be non-zero, but
if they are small, one can develop a series expansion in powers of $g J/M^2$ for the exact
solution.  In classical field theory, such expansions often have finite radius of convergence.

The effective action approach is simpler and in many
ways more physical, as it focuses attention on the relevant variables.  As outlined at
the end of \S \ref{sss:mag}, one can partially
understand the no-go theorem given here in these terms --
both flux and negative curvature sources of energy fall off too rapidly at large volume
to allow a local minimum.\footnote{
One can also understand time-dependent solutions with negative curvature, such as 
\cite{Townsend:2003fx}, in these terms -- kinetic energy allows rolling up the potential.
\cite{Emparan:2003gg}}

This interpretation of the no-go theorem suggests that one can get around it by incorporating
additional stringy effects in the potential.
While this may be, the need to satisfy the Einstein equation at every point in the internal
manifold suggests potential pitfalls with this.
One is that one can leave out a field which becomes
tachyonic at the solution of interest.  In the absence of arguments such as supersymmetry,
this could become more likely as one moves away from
a known higher dimensional solution.

A second potential pitfall is that the actual vacuum with massive modes may be qualitatively
different in some way from the original near-solution.  In the present case of constructing
de Sitter, there are at least two ways this could happen:
\begin{enumerate}
\item As one turns on massive fields, the value of the potential will decrease.  This
follows from standard variational arguments, or could be thought of as following from
a ``flow'' down the potential gradient.  Thus, even if the original value of the potential
is positive, the value at the true solution could be negative.
\item Symmetries or near-symmetries of the near-solution might be broken in this process.
For example, a shift symmetry which would tell us that the effective potential is flat in some
direction, might be broken.
\end{enumerate}
In fact, because of the argument we just gave
that one needs the warp and/or conformal factors to construct a negative curvature solution,
both possibilities would seem potentially relevant.  Any nontrivial warp or conformal factor
will by definition break translational symmetries on $M$.  And, the effective potential is
unbounded below under generic variations of these factors not satisfying the constraints.
Even with the constraint, there may be a lot of room to decrease the potential.

It seems very important to find some higher dimensional solutions, at least as explicit
as \cite{Giddings:2001yu} gave for IIb flux vacua, to evaluate these points.

\section*{Acknowledgments}

We thank  E.~ Bergshoeff, X. X.~ Chen, M.~ Hertzberg, S.~ Kachru, A.~ Linde, 
D.~ Lust, B.~ McInnes, G.~ Moore, D.~ Roest,
 E.~ Silverstein, G.~Shiu,  A.~ Van Proeyen and A.~ Westphal  for  useful discussions.  
We also thank I. Arefeva, J. Russo and P. Townsend for stressing to us that the arguments
here apply only to maximally symmetric space-times.
The work of MRD is supported by DOE grant DE-FG02-92ER40697, and 
the work of RK is supported by NSF grant 0756174.

\appendix

\section{Type IIA,  IIB $d=10$ Supergravity Energy-Momentum Tensor}

Here we review the democratic version of $d=10$ supergravity \cite{Bergshoeff:2001pv},
in which it is manifest that the stress-tensors for form fields is quadratic.  It also has the
virtue of treating the IIA, IIB, and massive IIA cases in a unified way.

The idea is to write a ``pseudo-action,'' which is a functional of both the
R-R  potentials and their duals.  To remove this doubling of the number of degrees of freedom,
one imposes the duality relations as additional constraints,
which do not follow from varying the pseudo-action.  The combined system of variational
equations and constraints is equivalent to the standard equations of motion
of the theory.

The pseudo-action has the extended field content
\begin{eqnarray}
{\rm IIA}&:&\hskip 1truecm  \left\{
  g_{\mu \nu},
  B_{\mu \nu},
  \phi,
  \C{1}_{\mu},
  \C{3}_{\mu \nu \rho},
  \C{5}_{\mu \cdots \rho},
  \C{7}_{\mu \cdots \rho},
  \C{9}_{\mu \cdots \rho},
  \psi_\mu,
  \lambda
  \right\} \, , \cr
{\rm IIB}&:&\hskip 1truecm  \left\{
  g_{\mu \nu},
  B_{\mu \nu},
  \phi,
  \C{0},
  \C{2}_{\mu \nu},
  \C{4}_{\mu \cdots \rho},
  \C{6}_{\mu \cdots \rho},
  \C{8}_{\mu \cdots \rho},
  \psi_\mu,
  \lambda
  \right\} \, .
\label{fcdemo}
\end{eqnarray}
It is understood that in the IIA case the fermions contain both
chiralities, while in the IIB case they satisfy
$
  \Gamma _{11} \psi_\mu = \psi_\mu\,,  \Gamma _{11} \lambda = -
  \lambda
$.
The bosonic part of the pseudo-action in Einstein frame is
\be
 S =
 - \frac{1}{2}\int d^{10} x \sqrt{-g}
    \Big[
    R\big(\omega(e)\big) +{1\over 2} \big( \partial{\phi} \big)^{2}
    +\tfrac{1}{2} e^{-\phi} H \cdot H\Big ] +  \sum_{n=0,1/2}^{5,9/2}
    \tfrac{1}{4} e^{(5-2n)\phi/2} \G{2n} \cdot \G{2n}
    \, .
    \label{IIABactiondemo}
\ee
The summation in the above pseudo-action is over
integers ($n=0,1,\ldots ,5$) in the massive IIA case,
over the integers ($n=1,\ldots ,4$) for standard IIA,
 and over half-integers
($n=1/2,3/2,\ldots ,9/2$) for IIB. In the summation range we will
always first indicate the lowest value for the IIA case, before the one
for the IIB case. For notational convenience we group all potentials and field
strengths in the formal sums
\begin{align}
  {\bf G} = \, \sum_{n=0,1/2}^{5,9/2} \G{2n} \,, \hspace{1cm}
  {\bf C} = \, \sum_{n=1,1/2}^{5,9/2} \C{2n-1} \,.  \label{formsums}
\end{align}
where $G^{p} \equiv {1\over p!}G_{M_1...M_{p}}dx^{M_1} \wedge ... \wedge  dx^{M_{p}}$ and $C^{q} \equiv {1\over q!}C_{M_1...M_{q}}dx^{M_1} \wedge ... \wedge  dx^{M_{q}}$.
Here the meaning of each term in the sum in the action is
\be
 \G{p} \cdot \G{p} \equiv {1\over p!} G_{M_1...M_{p}} G_{N_1...N_{p}} g^{M_1 N_1}...g^{M_p N_{p}}
\ee
The bosonic field strengths are given by
\begin{align}
  H = d B \,, \qquad
  {\bf G} = d {\bf C} - d B \wdg {\bf C} + \G{0} {\bf e}^B \,,
\label{G2n}
\end{align}
where it is understood that each equation involves only one term from the
formal sums (\ref{formsums}) (only the relevant combinations are
extracted). The corresponding Bianchi identities then read
$
  d H =0 \,,
  d {\bf G} - H \wdg {\bf G} =0 \,.
$
Here $G^{(0)} =m$ is the constant mass parameter of
IIA supergravity. In the IIB theory all equations should be read with
vanishing $\G{0}$.

Due to the appearance of all R-R potentials, the number of degrees of
freedom in the R-R sector has been doubled. 
To get the correct number of
degrees of freedom, we impose the  duality
relations in the bosonic sector of the theory which in the Einstein frame reads
\begin{align}
G^{(2n)} = (-)^{\Int[n]} e^{(2n-5)\phi/2} {\star G^{(10-2n)}}\, ,
\label{eq:duality}
\end{align}
These duality relations are combined with the equations of motion that follow from the pseudo-action
(\ref{IIABactiondemo}).

Our purpose here is to identify the energy-momentum tensor for all $d=10$ supergravity fields which in the Einstein frame defines the right hand side of the Einstein tensor.  For the Einstein equations the only thing that matters is the tensor character of the flux, how many contractions $g^{MN}$ are there. The fact that the flux is multiplied by a function of the dilaton does not matter. It matters that the terms  proportional to $(\partial \phi)^2$, $H \cdot H$ and $ \G{2n} \cdot \G{2n}$ all have the same sign as one can see from the action (\ref{IIABactiondemo}).  This means that the contribution to $T_{MN}$ from all $d=10$ supergravities comes from a class of energy-momentum tensors quadratic in antisymmetric $p$-rank tensors $F_{M_1,...M_p}$ where $p=0, ..,9$ which all have positive sign in the Einstein frame. This forms a basis for the detailed computations in section 2.


\end{document}